\documentclass{aastex}
\usepackage{emulateapj5}

\begin{document}
	\slugcomment{Accepted by AJ}

\title{Mapping the Inner Halo of the Galaxy with 2MASS-Selected
Horizontal-Branch Candidates}

\author{Warren R.\ Brown,
	Margaret J.\ Geller,
	Scott J.\ Kenyon}

\affil{Smithsonian Astrophysical Observatory, 60 Garden St, Cambridge, MA
02138}

\author{Timothy C.\ Beers}

\affil{Department of Physics and Astronomy, Michigan State University,
East Lansing, MI 48824}

\and
\author{Michael J.\ Kurtz,
        John B.\ Roll}

\affil{Smithsonian Astrophysical Observatory, 60 Garden St, Cambridge, MA
02138}

\shorttitle{Mapping the Galactic Halo with 2MASS-selected BHB Candidates}
\shortauthors{Brown et al.}

\begin{abstract}

	We use 2MASS photometry to select blue horizontal-branch (BHB)
candidates covering the sky $|b|>15^{\circ}$. A $12.5<J_0<15.5$ sample of
BHB stars traces the thick disk and inner halo to $d_{\sun}<9$ kpc, with a
density comparable to that of M giant stars. We base our sample selection
strategy on the Century Survey Galactic Halo Project, a survey that
provides a complete, spectroscopically-identified sample of blue stars to
a similar depth as the 2MASS catalog. We show that a $-0.20<(J-H)_0<0.10$,
$-0.10 < (H-K)_0 < 0.10$ color-selected sample of stars is 65\% complete
for BHB stars, and is composed of 47\% BHB stars. We apply this
photometric selection to the full 2MASS catalog and see no spatial
overdensities of BHB candidates at high Galactic latitude,
$|b|>50^{\circ}$. We insert simulated star streams into the data and
conclude that the high Galactic latitude BHB candidates are consistent
with having no $\sim$$5^{\circ}$ wide star stream with density greater
than 0.33 objects deg$^{-2}$ at the 95\% confidence level. The absence of
observed structure suggests there have been no major accretion events in
the inner halo in the last few Gyrs. However, at low Galactic latitudes a
two-point angular correlation analysis reveals structure on angular scales
$\theta\lesssim1^{\circ}$. This structure is apparently associated with
stars in the thick disk, and has a physical scale of 10-100 pc.
Interestingly, such structures are expected by cosmological simulations
that predict the majority of the thick disk may arise from accretion and
disruption of satellite mergers.

\end{abstract}

\keywords{
        stars: early types ---
        Galaxy: stellar content ---
        Galaxy: halo}

\section{INTRODUCTION}

	The formation and evolution of the Milky Way galaxy remains one of
the outstanding questions of modern astronomy. Recent observations and
n-body simulations lend increasing support to a hierarchical assembly
model, where the halo of our Galaxy is composed (at least in part) of
tidally disrupted dwarf galaxies. N-body models suggest that dwarf
galaxies disrupted long ago should still be visible as coherent streams of
stars within the Galactic halo
\citep{johnston95,johnston96,helmi99b,harding01,bullock01}. The most
striking example is the discovery of the Sagittarius dwarf galaxy in the
process of being tidally disrupted by the Milky Way \citep{ibata94}.
\citet{majewski03} detect stellar debris from the Sgr dwarf over much of
the sky. At least one additional stream, the so-called Monoceros stream,
surrounds the disk of the Milky Way, and may be associated with another
disrupted satellite galaxy \citep{newberg02,yanny03,ibata03}. Clearly,
examination of the locations, motions, and compositions of the stars in
the halo (and thick disk) should provide us with a more complete record of
the Milky Way's formation history.

	Previous surveys have demonstrated that blue horizontal branch
(BHB) stars provide excellent tracers of the stellar halo \citep{pier82,
sommer89, preston91, arnold92, kinman94, wilhelm99b, brown03}. BHB stars
are numerous, exceeding the number density of RR Lyraes by roughly a
factor of 10 \citep{preston91}. BHB stars are also luminous, and hence
observable to large distances. BHB stars exhibit a small dispersion in
absolute magnitude, making reasonably accurate photometric distance
estimates possible. Furthermore, BHB stars are bluer than most competing
stellar populations, making their identification on the basis of broadband
colors relatively straightforward.

	In the past, objective-prism surveys were the primary source of
candidate BHB stars \citep{sommer86,preston91b,arnold92,beers96}, often
supplemented by $UBV$ and Str\"{o}mgren photometry. The recent work of
\citet{ivezic00}, \citet{yanny00}, \citet{vivas01}, \citet{newberg02},
\citet{vivas03}, and \citet{newberg03} shows that photometric surveys can
be used to identify structure in number counts of A-type, F-type, and RR
Lyrae stars at distances of up to $\sim$100 kpc.

	The Two Micron All Sky Survey \citep[2MASS,][]{cutri03} now
provides complete, uniform $JHK$ photometry over the entire sky. Here we
demonstrate that two-color near-infrared photometry can also be used to
efficiently select candidate BHB stars. A properly selected set of 2MASS
BHB candidates will permit, for the first time, an all-sky survey of the
``inner'' Galactic halo. BHB stars at the limiting apparent magnitude of
the 2MASS catalog ($J<16$) sample the halo of the Galaxy up to
helio-centric distances of $d_{\sun}<11$ kpc. This corresponds to a
maximum Galacto-centric distance of $r_{GC}\sim 20$ kpc in the anti-center
direction.

	We study number counts of objects with BHB colors in the 2MASS
catalog, and find no obvious overdensities at high Galactic latitudes that
might be associated with known or newly identified streams. This lack of
projected spatial structure emphasizes the need to obtain full
six-dimensional kinematic information provided by radial velocities and
proper motions. Conveniently, 2MASS-selected BHB stars are bright enough
to be included in the existing UCAC2 \citep{zacharias00} and SPM 3.0
\citep{girard03} proper motion catalogs. In addition, $J<16$ stars can be
observed with moderate signal-to-noise spectroscopy on 1m-2.5m class
telescopes.

	The catalog of BHB candidates we provide herein forms the basis
for a uniform spectroscopic survey. A spectroscopic survey of
2MASS-selected BHB candidates is particularly well-suited to study the
structure of the inner halo and thick disk. Sufficiently accurate radial
velocities and proper motions will permit identification of star streams
at small scales, in particular through inspection of angular momentum
phase space \citep{helmi99,chiba00,helmi03}, and for measurement of the
global rotation of the inner halo at the largest scales. Previous surveys
have found evidence for (i) no halo rotation \citep{layden96,
gould98,martin98}, (ii) a small prograde rotation \citep{chiba00}, and
(iii) retrograde rotation \citep{majewski92, spagna02}. Perhaps these
conflicting observational results indicate that the halo velocity field
has substructure, an issue best addressed by an all-sky survey. The
important question of how the metal-weak thick disk \citep{beers02} is
kinematically related to the inner halo population can also be pursued
with such a survey.

	We begin by studying the efficacy of using 2MASS near-IR
photometry to select BHB stars. In \S 2 we introduce the Century Survey
Galactic Halo Project, a survey that provides a complete,
spectroscopically-identified sample of blue stars to the depth of the
2MASS photometry. In \S 3 we consider the ability of 2MASS photometry to
differentiate BHB stars from other blue objects, including A-type stars of
higher surface gravity (many of which are likely halo and thick-disk blue
stragglers) in the Century Survey Galactic Halo Project sample. In \S 4 we
apply a two-color photometric selection to the full 2MASS catalog, and
look for overdensities in the number counts of objects with BHB-like
colors. In \S 5 we discuss a two-point angular correlation analysis of the
2MASS-selected objects, and use simulated star streams to understand our
sensitivity to structure.  We conclude in \S 6.

\section{THE CENTURY SURVEY GALACTIC HALO PROJECT SAMPLE}

	The Century Survey Galactic Halo Project is a photometric and
spectroscopic survey from which we select relatively blue stars as probes
of the Milky Way halo. \citet{brown03} includes a detailed description of
the sample selection, data reduction, and analysis techniques. In brief,
we obtained Johnson $V$ and Cousins $R$ broadband imaging for a
$1^\circ\times64^\circ$ strip using the 8 CCD MOSAIC camera
\citep{muller98} on the KPNO 0.9 m telescope. The average depth of the
photometry is $V=20.3$. We then use the CCD photometry to select blue
$(\vr) <0.30$ stars with $V<16.5$ for follow-up spectroscopy using the
FAST spectrograph \citep{fabricant98} on the Whipple Observatory 1.5m
telescope. Moderate signal-to-noise (S/N$\approx$30), medium-resolution
(2.3 \AA) spectra allow us to measure radial velocities, temperatures,
surface gravities, metallicities, and spectral types for the stars.

	The Century Survey Galactic Halo Project sample contains 764
objects. In this paper we make use of the 553 objects with $V<16$. We
choose the $V<16$ cut to match the depth of the 2MASS catalog and to avoid
objects with poor near-IR photometry. The Century Survey Galactic Halo
Project sample consists predominantly of F- and A-type stars plus a small
number of unusual objects (i.e., white dwarfs and subdwarfs). The A-type
stars have a large range of metallicity ($-3<$ [Fe/H] $<0$), a large
velocity dispersion ($\sigma = 98$ km s$^{-1}$), and distance estimates
that identify them as members of the inner halo and thick-disk populations
\citep{brown03}.

	One of the primary goals of the Century Survey Galactic Halo
Project is to use BHB stars to trace potential star streams in the halo.
The primary difficulty in using BHB stars as tracer objects is the need to
distinguish reliably between low surface-gravity BHB stars and the higher
surface-gravity A dwarfs and blue stragglers. In \citet{brown03} we devote
careful attention to the reliable classification of BHB stars. We apply
the techniques of \citet{kinman94}, \citet{wilhelm99a}, and
\citet{clewley02}, and find 26 high likelihood BHB stars among the 96
A-type stars with $V<16$. We use this spectroscopically identified sample
of stars to test the efficacy of 2MASS photometry for selecting BHB stars.

	Figure \ref{fig:vrsample} shows the (\vr)$_0$ distribution of the
$V<16$ stars in the Century Survey Galactic Halo Project. All of the BHB
stars fall within the color range $-0.15 < (\vr)_0 < 0.10$. Figure
\ref{fig:vrsample}b, a useful guide for observers, plots the fraction of
A-type and BHB stars found in samples selected by $(\vr)_0$ less than the
color marked on the x-axis. BHB stars, for example, constitute 54\% of all
($V<16$) stars selected with $(\vr)_0 < 0.10$. Thus, broadband color
selection is an efficient selection criteria for BHB stars, and one that
we explore for the 2MASS catalog.

\section{EFFICACY OF 2MASS PHOTOMETRIC SELECTION}

	We access the complete 2MASS point source catalog
\citep{cutri03}\footnote{Available at
\url{http://www.ipac.caltech.edu/2mass/}.} and find matches for every
object in the Century Survey Galactic Halo Project. The average 2MASS
photometric uncertainty in the colors of the $V<16$ Century Survey
Galactic Halo Project objects is $\sigma(J-H)=\pm0.04$ and $\sigma(H-K)
=\pm0.05$. Four objects have atypically poor errors $\sigma(J-H)>0.11$; we
exclude these stars from the analysis. The $V<16$ Century Survey Galactic
Halo Project sample has a $J$-band magnitude limit of $J<15.5$.

	Figure \ref{fig:jhsample} shows the 2MASS $(J-H)_0$ and $(H-K)_0$
color distribution of the $V<16$ Century Survey Galactic Halo Project
sample from Figure \ref{fig:vrsample}. It is clear from 

 \includegraphics[width=6.5in]{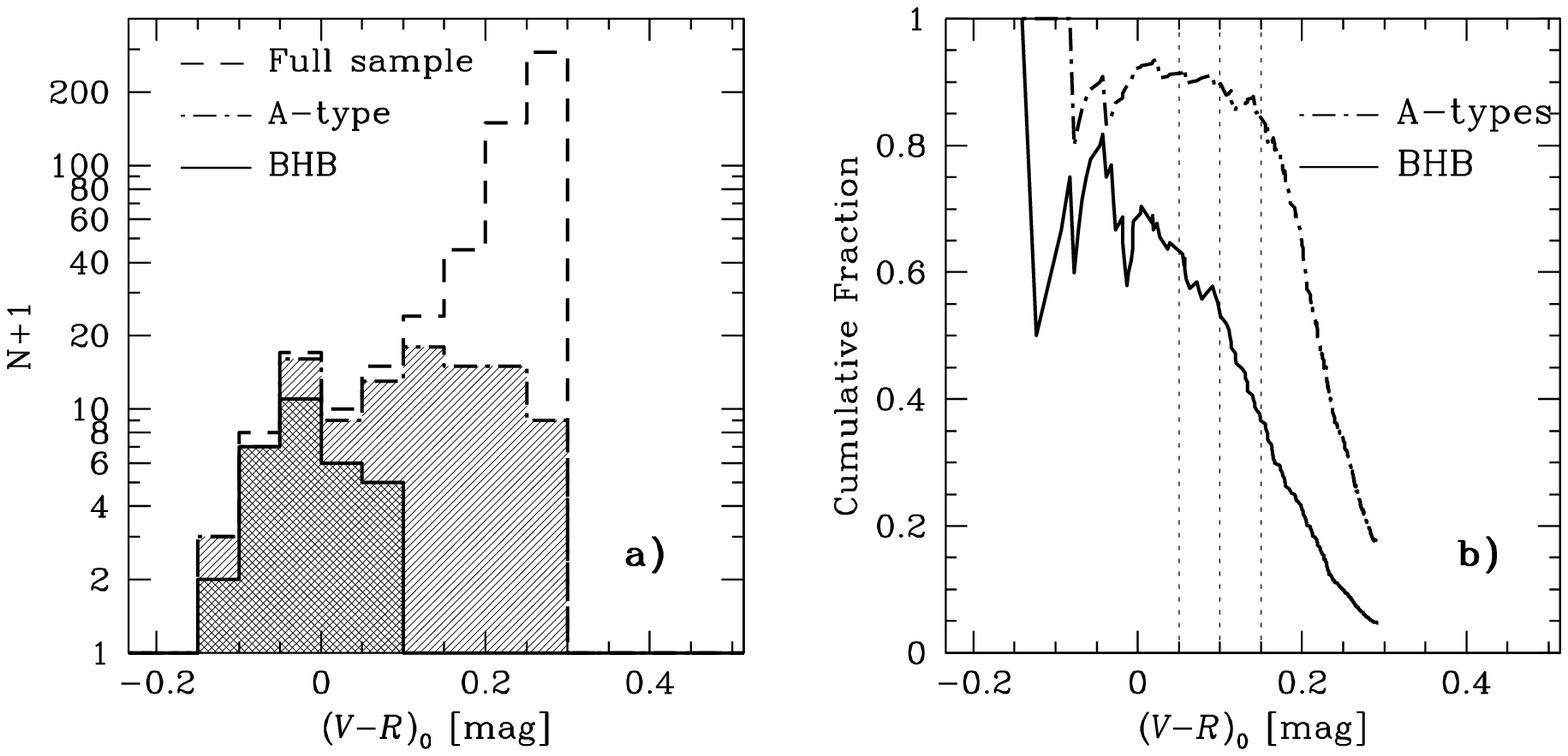}
 \figcaption{ \label{fig:vrsample}
	Distribution in $(V-R)_0$ for 553 stars from the Century Survey
Galactic Halo Project selected with $V<16$.  Panel (a) shows the
histogram of the full sample, A-type stars, and BHB stars. Panel (b)
shows the fraction of A-type (dot-dashed line) and BHB (solid line) stars
selected with a $(\vr)_0$ bluer than the indicated color; the vertical
dotted lines at colors of 0.05, 0.10, and 0.15 are provided to help guide
the eye.}

	~

	\noindent Figure \ref{fig:jhsample}a that $(J-H)_0$ provides a
useful discriminant between the F-type, A-type, and BHB stars in the
Century Survey Galactic Halo Project sample. The BHB stars are
substantially bluer than the majority of competing objects. By contrast,
$(H-K)_0$ (Figure \ref{fig:jhsample}b) provides much less discrimination
between the F-type, A-type, and BHB stars.

	Figure \ref{fig:jhsample}c shows that BHB stars constitute 41\% of
stars selected with colors $(J-H)_0<0.10$, comparable to the (\vr) sample
selection.  A $(J-H)_0<0.10$ sample of stars is 65\% complete for BHB
stars.  A $(J-H)_0<0.15$ selected sample of stars, on the other hand, is
96\% complete for BHB stars, but BHB stars constitute only 29\% of the
sample.  We conclude that a $(J-H)_0<0.15$ selection is optimal for
completeness; a $(J-H)_0<0.10$ selection is optimal for observational
efficiency.  We employ the $(J-H)_0<0.10$ color selection criteria in the
following sections.

	We can use the $(H-K)_0$ color to exclude clear {\it non}-BHB
stars from our samples. Figure \ref{fig:jhsample}b shows that all BHB
stars fall within the color range $-0.10 < (H-K)_0 < 0.10$. Applying this
$(H-K)_0$ color limit to the $(J-H)_0<0.10$ and $(J-H)_0<0.15$ samples
improve their identification efficiency to 47\% and 32\%, respectively. We
use the $-0.10 < (H-K)_0 < 0.10$ color selection criteria in the following
sections.

\section{2MASS-SELECTED ALL-SKY MAPS OF BHB CANDIDATES} \label{sec:maps}

	We use the two-color selection discussed above to generate all-sky
maps from the complete 2MASS point source catalog. Our present goal is to
look for obvious projected spatial structure in the distribution of BHB
and A-type stars. \citet{majewski03} perform a similar analysis, selecting
M giants from the 2MASS point catalog, and find dramatic tidal streams
from the Sagittarius dwarf galaxy 

	~

	\vskip 3.8in

	~

\noindent circling the sky. The major difference between our maps and
those of \citet{majewski03} is that the absolute magnitude of a BHB star
is fainter than for an M giant. However, we expect that the density of BHB
stars is comparable to the density M giants.

	Given the reported detection of Sgr-stream M giants, it is
informative to quantify the number of BHB stars that might be detected
with sufficiently deep samples. Deriving a ratio of BHBs to M giants is
complicated by the fact that M giants are found in metal-rich populations,
while BHB stars are found largely in metal-poor populations. Tidal streams
from a dwarf galaxy merger will likely possess {\it both} metal-rich and
metal-poor populations. The Sgr stream, for example, has been identified
both with giants \citep{majewski03,ibata02,kundu02,ibata01b} and with
horizontal branch stars \citep{monaco03,vivas01,yanny00,ivezic00}.  For
this exercise, we assume that stars in a tidal stream produce horizontal
branch (HB) stars and red giant (RG) stars at an equal rate, and estimate
the ratio of HB stars to RG stars by inspecting their lifetimes on
Yonsei-Yale isochrones \citep{yi01}.  In particular, we consider the ratio
of bright giants near the tip of the RG branch to the HB with the
isochrones populated by giants with luminosities brighter that the Yale
horizontal-branch location, i.e., $L/L_{\sun} \ge 50 L_{\sun}$.


	According to the Yonsei-Yale isochrones, and the work of
\citet{yong00}, the lifetime of a star on the HB is 100-150 Myr. This
timescale is almost independent of helium content, for Y = 0.24-0.29, and
metallicity, Z, for Z = 0.001 to Z = 0.02. We assume that the majority of
HB lifetime is spent at roughly constant luminosity before the star
evolves up onto the AGB (if the envelope mass is large) or onto a white
dwarf cooling curve (if the envelope mass is small).

 \includegraphics[width=6.5in]{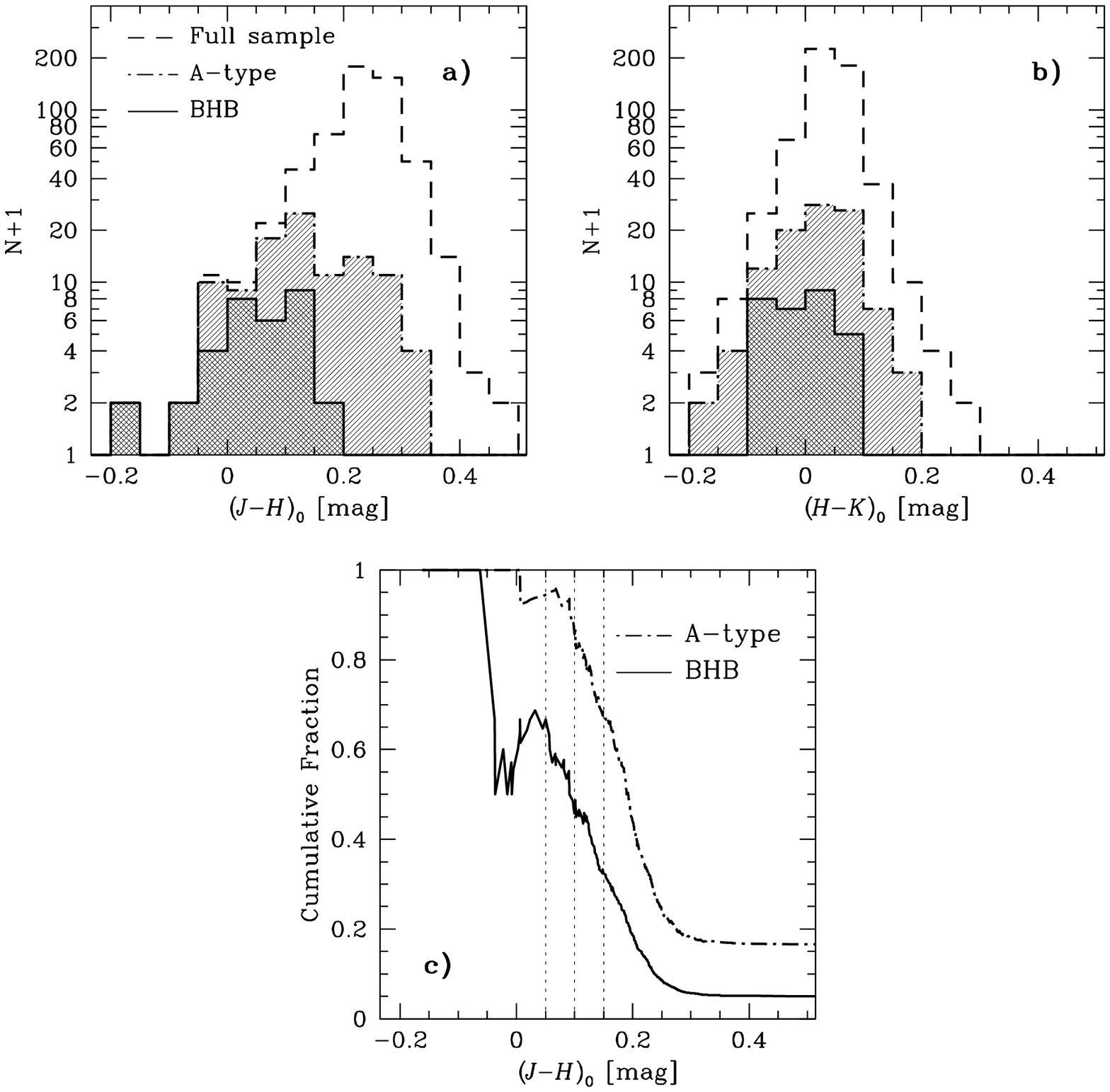}
 \figcaption{ \label{fig:jhsample}
	2MASS colors of the $V<16$ Century Survey Galactic Halo
Project sample.  Panels (a) and (b) show the $(J-H)_0$ and $(H-K)_0$
histograms, respectively, of the full sample, the A-type stars, and the
BHB stars. Panel (c) shows the fraction of A-type and BHB stars selected
with $(J-H)_0$ bluer than the indicated color; the vertical dotted lines
at colors of 0.05, 0.10, and 0.15 are intended to help guide the eye.}

	~

	For the bright RG stars, the lifetime spent with $L/L_{\sun} >
100$ is 30 Myr for Z = 0.01-0.02 and 35-40 Myr for Z = 0.001, for stars
with masses in the range $1 \le M/M_{\sun} \le 2$. The lifetime for
$L/L_{\sun} > 50$ is 70-80 Myr, and is again roughly independent of Z. The
lifetime for $L/L_{\sun} > 200$ is 15-20 Myr. The most luminous (massive)
stars are not expected to contribute significantly in old stellar
populations. Lower mass stars spend very short times near the tip of the
RG branch, hence will also have little impact on our rough estimates.

	Comparing the above lifetimes, we estimate the relative number of
HB:RG stars in a volume-limited sample should be on the order 5:1 for
$L/L_{\sun} > 200$ red giants, 3:1 for $L/L_{\sun} > 100$ red giants, and
1.5-2.0:1 for $L/L_{\sun} > 30$ 

	~

\vskip 7.1in

	~

\noindent red giants. The \citet{majewski03} selection criteria and survey
depth suggests that they trace $L > 300 L_{\sun}$ M giants, though they
should also detect $L$ = 100-200 $L_{\sun}$ red giants to 10 kpc. Thus the
2MASS BHB sample should include 3-5 HB stars for every M giant star in the
\citet{majewski03} maps.

	Because the Yonsei-Yale isochrones are unable to distinguish BHB
stars from other HB stars, the HB:RG ratio 3-5:1 is likely an upper limit
to the BHB:RG ratio.  The BHB:RG ratio ultimately depends on the
metallicity of the stellar population:  color-magnitude diagrams of metal
rich globular clusters, for example, show that the red giant branch is far
more populated than the horizontal branch \citep[e.g.][]{rosenberg00}.  
\citet{monaco03} study BHB stars in the Sgr stream and find that a
metal-poor population constitutes $\sim$10\% of the stellar population.  
This implies that the BHB:RG ratio in the predominantly metal-rich Sgr
stream is more like 1:2-3.  We conclude that the number density of BHB
stars is comparable, on average, to M giants in tidal debris from
satellites like the Sgr dwarf.  It is not known whether the halo is
composed entirely of tidal debris; it is likely that the number density of
BHB stars exceeds that of M giants in the metal poor halo.

	We now map the inner halo with BHB and A-type stars. We begin with
the full 2MASS point source catalog of 470,992,970 objects. We select
objects that have photometric quality flags of A or B (objects with
photometric errors less than $\pm 0.155$ mag), and reject objects with
photometric quality flags D, E, F, U, or X (objects with poor or
non-existent photometry). We also reject objects with contamination flags
p, d, s, and b (objects contaminated by nearby bright stars or other
objects, as well as CCD artifacts). Finally, we reject all objects in the
region of the Galactic plane $-15^{\circ} < b < +15^{\circ}$. Our interest
is in viewing the halo uncontaminated by thin-disk stars and heavy
reddening from dust and gas in the Galactic plane. The Galactic latitude
selection reduces the 2MASS point source catalog to a more manageable
78,464,293 objects. We perform a preliminary color cut $(J-H)<0.3$ and
$(H-K)<0.5$ to further reduce the catalog size to 7,426,805 blue objects.
We then calculate and apply reddening corrections from \citet{schlegel98}
to derive intrinsic de-reddened colors $(J-H)_0$ and $(H-K)_0$.

	Our working catalog of BHB candidates contains 99,431 2MASS point
sources selected {\it after} reddening corrections by $12.5 < J_0 < 15.5$,
$-0.2 < (J-H)_0 < 0.1$, and $-0.1 < (H-K)_0 < 0.1$. Based on our
comparison with the Century Survey Galactic Halo project, we expect that
47\% of the these blue point sources are BHB stars, 39\% are higher
gravity A-type stars, and 14\% are miscellaneous objects (mostly early
F-types).

	Figure \ref{fig:skymaps} shows equal-area Hammer-Aitoff
projections of the two-color selected 2MASS BHB candidates. We show three
magnitude ranges. The top panel shows the apparent magnitude range
$12.5<J_0<13.5$, the middle panel $13.5<J_0<14.5$, and the bottom panel
$14.5<J_0<15.5$. We bin objects into pixels of area 1 square degree. The
sky maps are in Galactic coordinates centered on
$(l,b)=(180^{\circ},0^{\circ})$.

	Inspection of the sky maps reveals a number of salient features.
The LMC and SMC are clearly visible in all three magnitude ranges,
presumably due to the presence of their bright O- and B-type stars. The
Galactic plane is also clearly visible, despite our efforts to eliminate
it. Gould's belt slants down from $l\approx90^{\circ}$ to
$l\approx180^{\circ}$. Small empty regions near the $b=\pm15^{\circ}$
boundaries result from a combination of large reddening corrections and
from our preliminary color cut. Perhaps the most impressive feature in the
maps is the Galactic bulge, which extends nearly to the Galactic poles:
the surface number density of objects towards the Galactic center
$-90^{\circ}<l<90^{\circ}$ converges to the surface number density of
objects towards the anti-center $90^{\circ}<l<270^{\circ}$ at Galactic
latitude $|b|\sim75^{\circ}$ (see Figure \ref{fig:density}).

	The sky maps contain no obvious structures like the Sagittarius
stream evident in the sky maps of \citet{majewski03}. The small clump of
objects below the Galactic plane near $l=170^{\circ}$ (Figure
\ref{fig:skymaps}, bottom panel) is coincident with a clump in the M giant
map \citep{majewski03}, but it is offset from the Sagittarius stream by
approximately 15$^{\circ}$, and is probably not associated with this
structure.

	We now look for structure in space, but caution that distance
estimates to the BHB candidates are rough estimates at best. In principle,
distances to BHB stars can be accurately determined if the metallicity is
known. However, published $M_V$-metallicity relations for horizontal
branch stars vary considerably, with values of the slope ranging between
0.15 \citep{carney92} and 0.30 \citep{sandage93} and values of the zero
point falling into two groups, $\sim$0.30 mag apart. We use the {\it
Hipparcos}-derived zero point,

 \begin{center}
  \includegraphics[width=3.5in]{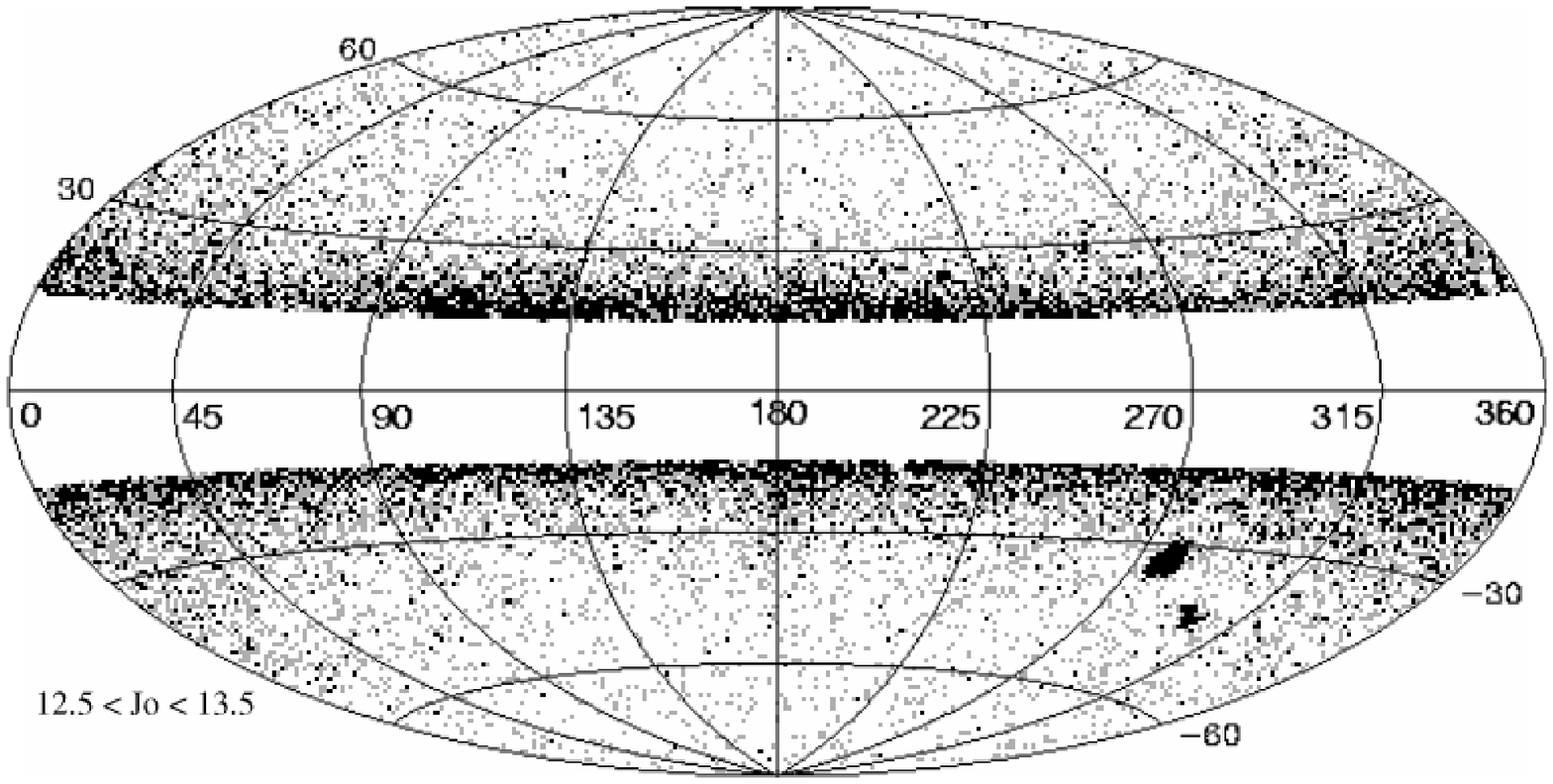}
  \includegraphics[width=3.5in]{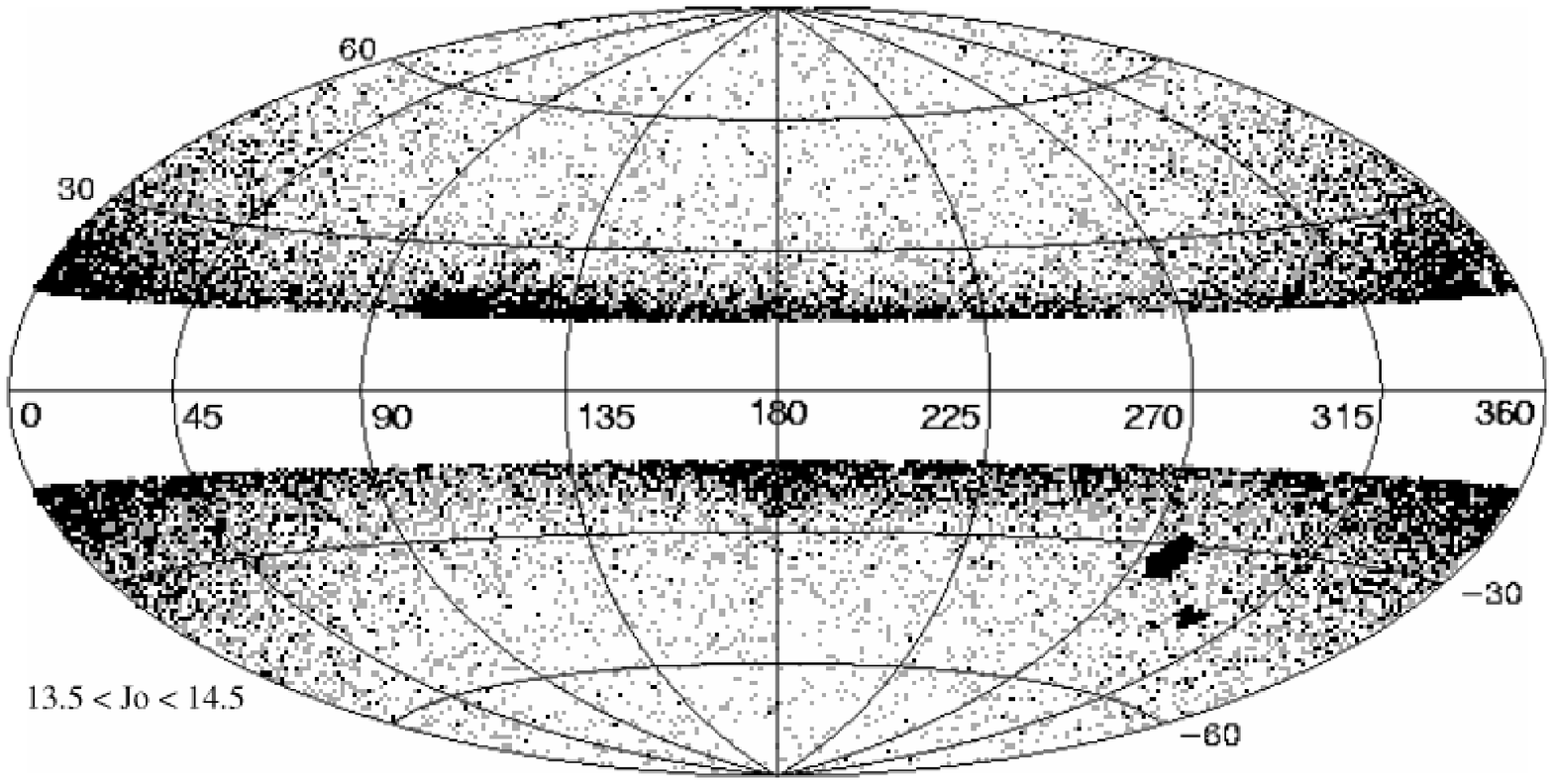}
  \includegraphics[width=3.5in]{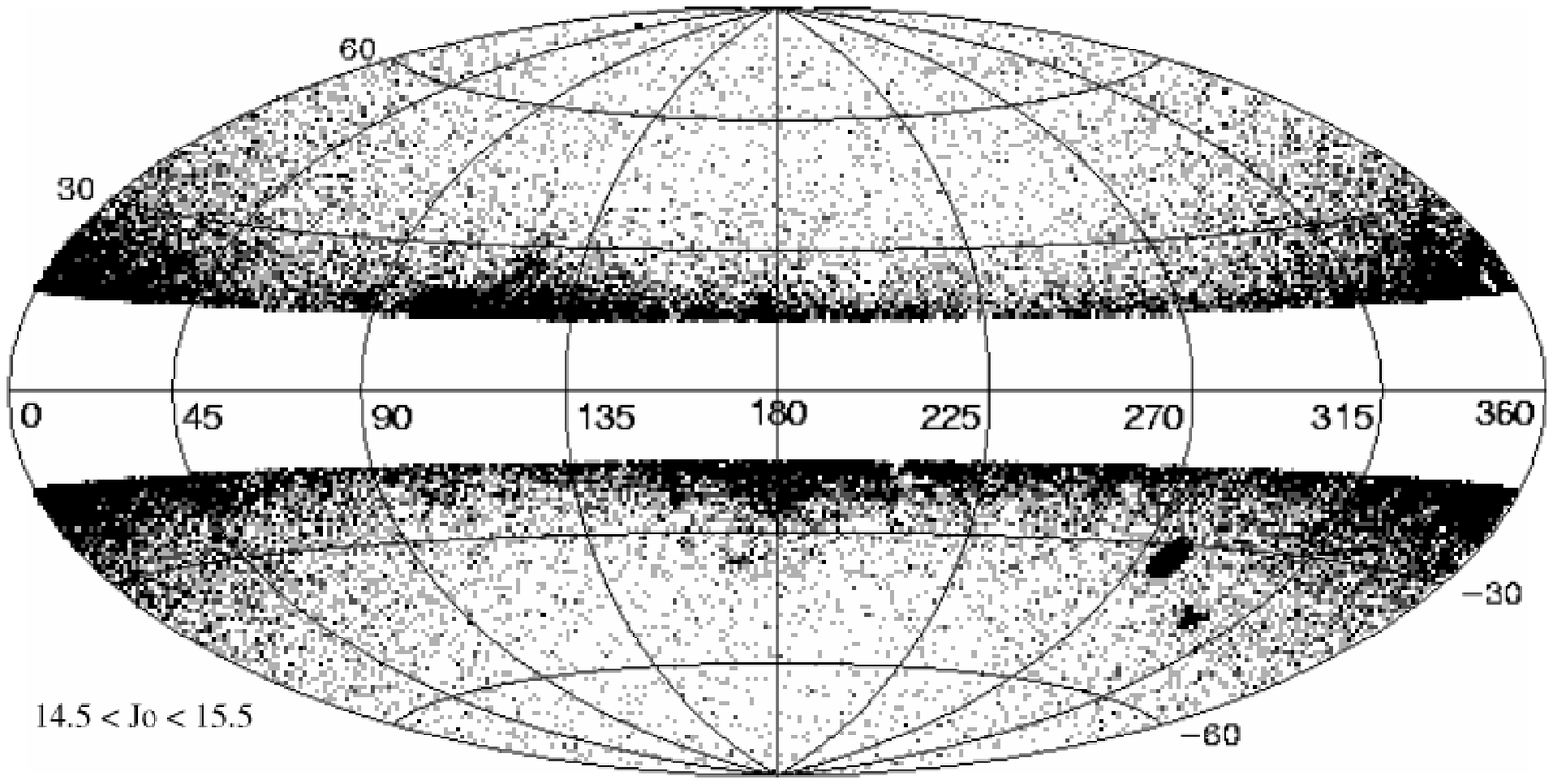}
 \end{center}
\figcaption{ \label{fig:skymaps}
        Sky maps in Galactic coordinates for the complete 2MASS point
source catalog selected by de-reddened color: $-0.2 < (J-H)_0<0.1$ and
$-0.1<(H-K) _0<0.1$. The top panel shows the apparent magnitude range
$12.5<J_0<13.5$, the middle panel $13.5<J_0<14.5$, and the bottom panel
$14.5<J_0<15.5$. Objects are binned into 1 deg$^2$ pixels. Comparison
with the Century Survey Galactic Halo Project suggests that the
color-selected objects are 47\% BHB stars, 39\% A-type stars, and 14\%
miscellaneous objects (mostly early F-types). Galactic latitudes
$-15^{\circ}<b<15^{\circ}$ have been removed for ease of viewing.}

	~

 \includegraphics[width=3.5in]{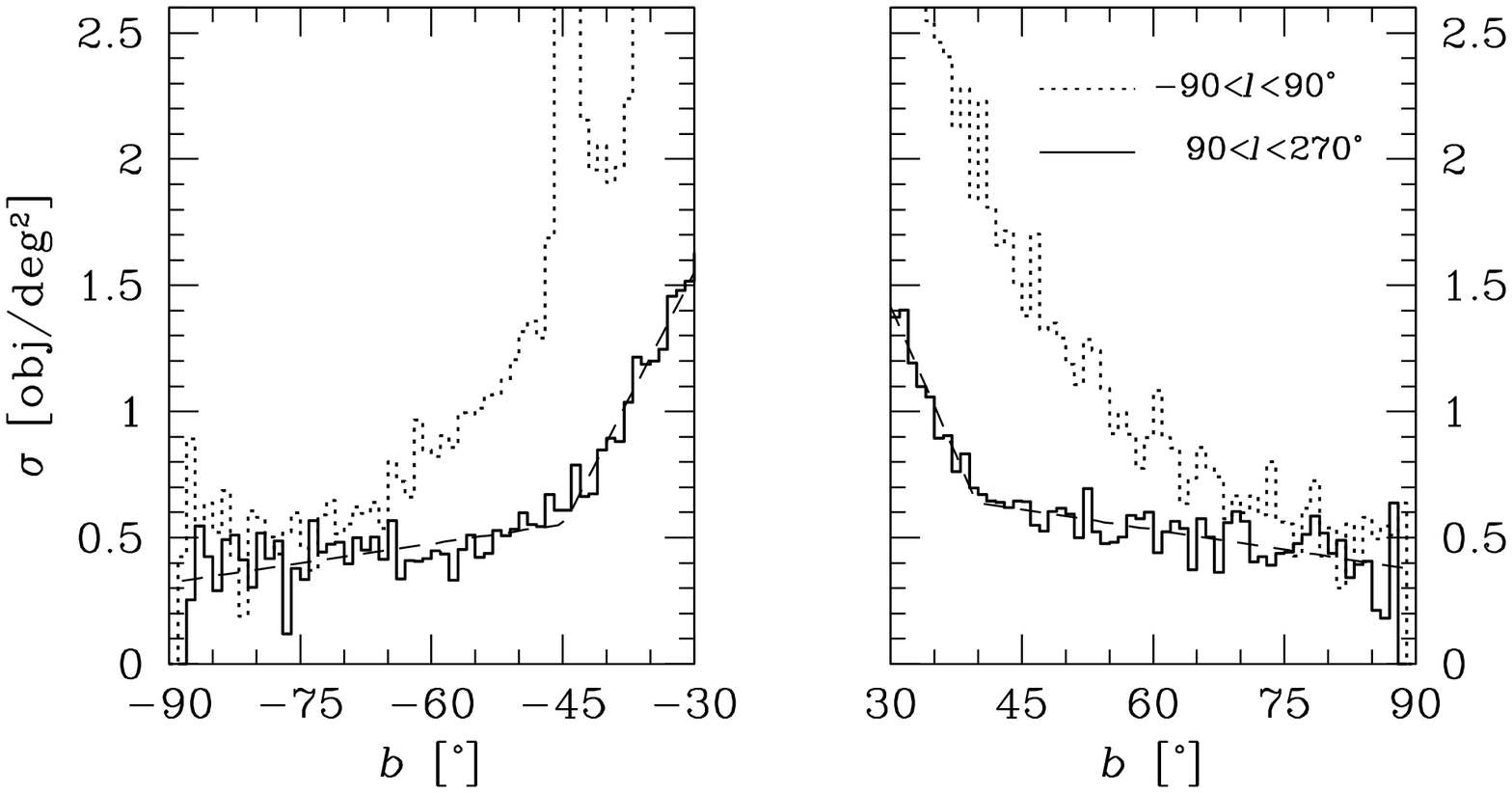}
 \figcaption{ \label{fig:density}
	The surface number density of $12.5 < J_0 < 15.5$ BHB candidates
selected $-90^{\circ} < l < 90^{\circ}$ (dotted line) and $90^{\circ} < l
< 270^{\circ}$ (solid line) as a function of Galactic latitude $b$. The
spike in number density at $b=-45^{\circ}$ is caused by the SMC.  Dashed
lines show the two-component fit that we use as the density distribution
for the two-point angular correlation random catalogs.}

	~

\noindent $M_V(RR) =0.77\pm0.13$ at [Fe/H] = $-$1.60
\citep{gould98}, based on the statistical parallax of 147 halo RR Lyrae
field stars. We employ the recently measured $M_V$-metallicity slope
$0.214\pm0.047$ \citep{clementini03}, based on photometry and spectroscopy
of 108 RR Lyrae stars in the Large Magellanic Cloud. We lack spectroscopic
metallicity determinations for the BHB candidates; we thus use the mean
metallicity of the Century Survey Galactic Halo Project BHB stars,
[Fe/H]=$-1.5$, to obtain $M_V=+0.80\pm0.14$. Because BHB stars have an
$\sim$A0 spectral type and, by definition, $(V-J)\simeq0.0$, this $M_V$
corresponds to $M_J\simeq+0.80$. BHB stars with more extreme metallicities
of [Fe/H]=0 or [Fe/H]=$-3$ will have absolute magnitudes that differ by
$\pm0.32$ -- a factor of $\pm$16\% in distance.

	For the remainder of this paper, we assume $M_J\simeq+0.80$
to obtain distance estimates to the 2MASS-selected BHB candidates.
The magnitude range $12.5< J_0 <15.5$ thus corresponds to an approximate
helio-centric distance range $2<d_{\sun}<9$ kpc. We note that this
distance range is substantially shorter for the 53\% of the BHB candidates
that

 \includegraphics[width=6.5in]{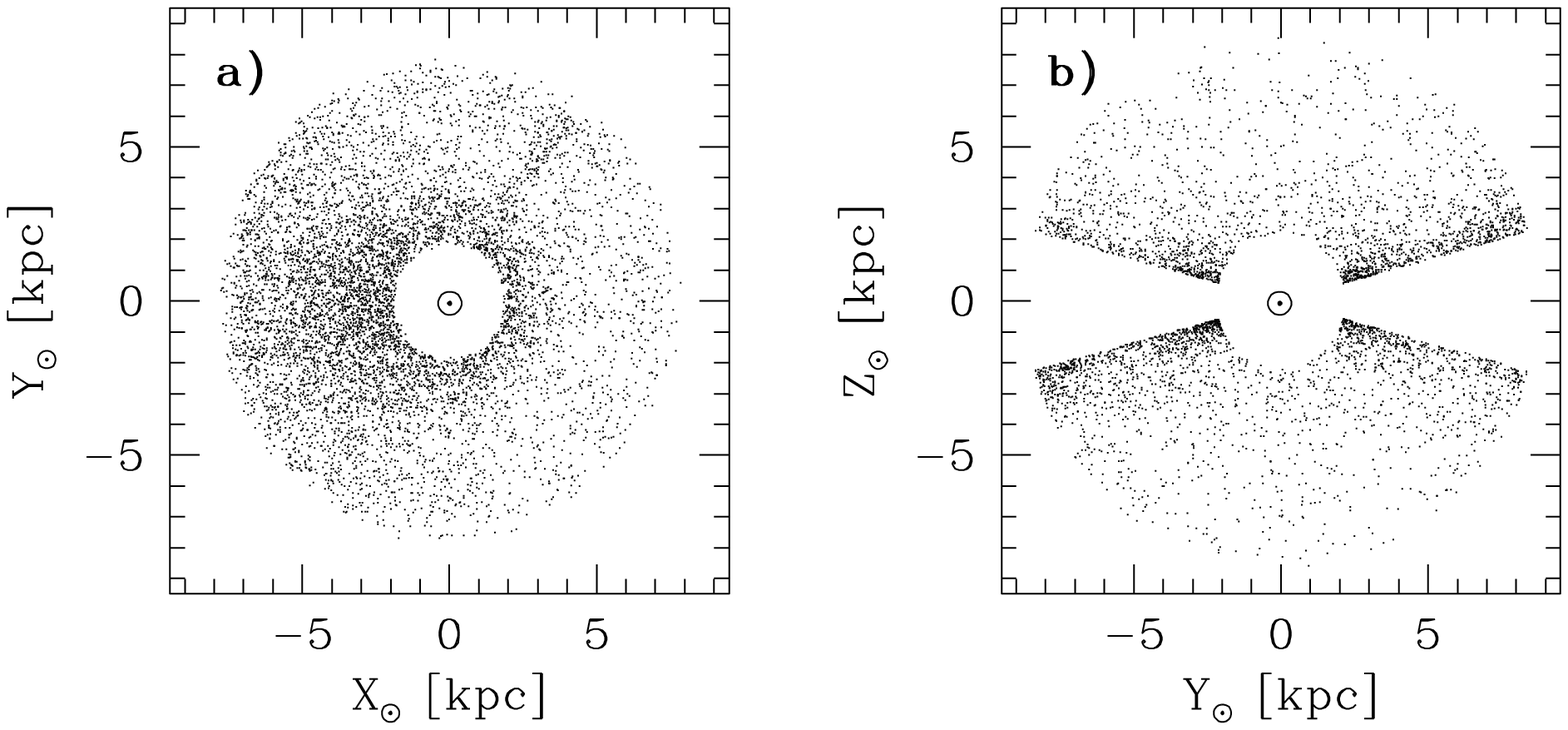}
 \figcaption{ \label{fig:slice}
Spatial distribution of 2MASS BHB candidates, assuming $M_J=+0.8$.
Panel a) is a $10^{\circ}$ wedge selected $25^{\circ} < b <
35^{\circ}$, and panel b) is a $10^{\circ}$ wedge through the Galactic
poles.  For reference, the Galactic center is located
$(X,Y,Z)=(-8.5,0,0)$ kpc.}

\noindent are A- or early F-type dwarfs. A5 and F0 dwarfs have $M_J=1.7$
and $M_J=2.2$, respectively \citep{allen00, bessell88}. The corresponding
distance ranges for A5 and F0 dwarfs are reduced by factors of 0.63 and
0.50, respectively, compared to BHB stars.

	We plot the distribution of BHB candidates in space (Figure
\ref{fig:slice}). Figure \ref{fig:slice}a shows the BHB candidates from a
$10^{\circ}$ wedge selected with $25^{\circ} < b < 35^{\circ}$. The
X$_{\odot}$ axis points in the direction of the Galactic anti-center,
$l=180^{\circ}$, the Y$_{\odot}$ axis points in the direction of solar
motion, $l=90^{\circ}$, and the Z$_{\odot}$ axis points in the direction
of the north Galactic pole, $b=+90^{\circ}$. In Figure \ref{fig:slice}a
the Galactic center is located at X$_{\odot}=-8.5$ kpc. Bulge objects
dominate the left-hand side of the plot. The radial feature matches the
Galactic plane structure at $l\sim120^{\circ}$ in Figure
\ref{fig:skymaps}. Figure \ref{fig:slice}b shows BHB candidates from a
$10^{\circ}$ wedge centered on the Y-Z plane that samples the Galactic
poles. The gap in the middle of Figure \ref{fig:slice}b is the Galactic
plane exclusion region. Although the 2MASS BHB candidates are not
distributed uniformly in space, we find no obvious spatial structure
except that associated with the Galactic plane.

	The distribution of inner-halo BHB candidates appears remarkably
smooth. Perhaps this uniformity is not surprising: existing observations
and simulations suggest that the inner halo {\it should} be smooth. Known
structures like the Sagittarius star stream and the Monoceros ring are
located at Galacto-centric distances greater than 20 kpc, beyond the
region sampled by the 2MASS-selected BHB candidates. In addition,
theoretical simulations suggest that star streams from disrupted
satellites become well-mixed with the underlying stellar populations of
the Galaxy after a few Gyrs, and show little spatial structure, especially
in the inner halo, where orbital time scales are relatively short
\citep{johnston96,helmi99b,bullock01}. This picture is in agreement with
\citet{gould03} who argues, based on the proper motions of
solar-neighborhood stars, that there are at least 400 streams in the local
stellar halo. On the other hand, \citet{helmi99} find evidence for inner
halo structure in angular 

	~

	\vskip 3.2in

	~

\noindent momentum space. This suggests that old star streams that lack
spatial coherence may still be detectable with six-dimensional
information.

\section{TWO-POINT ANGULAR CORRELATION FUNCTION OF BHB CANDIDATES}

	The two-point angular correlation function provides a quantitative
measure of structure, and is commonly used as a measure of large scale
structure in galaxy redshift surveys. Here, we apply the two-point angular
correlation function to the catalog of 2MASS-selected BHB candidates.
\citet{doinidis89} performed a similar two-point angular correlation
function analysis on 4,400 BHB candidates from the HK objective-prism
survey, and found an excess of stellar pairs with angular separations less
than 10 arcmin. Their result is significant at the 5-$\sigma$ level, and
motivates us to look for similar correlation in the 2MASS-selected BHB
candidates. Recently, \citet{lemon03} calculate angular correlation
functions for faint, F-colored stars located at $41^{\circ} < b <
63^{\circ}$ in the Millennium Galaxy Catalog. They find no signal at
angular scales less than $5^{\circ}$.

	We calculate the two-point angular correlation function using the
Monte Carlo estimator \begin{equation}
	\omega(\theta) = \frac{N_p(\theta)}{N_r(\theta)} - 1,
\end{equation} where $N_p(\theta)$ is the number of pairs in the data
catalog with separations in the range $\theta \pm \Delta\theta/2$, and
$N_r(\theta)$ is the number of pairs in each of the random catalogs. Using
the Monte Carlo method eliminates the need to calculate edge corrections
\citep{hewett82}.  We are concerned, as pointed out in \citet{doinidis89},
about the inclusion of BHB stars associated with globular clusters.
Correlated BHB stars from globular clusters could produce a spurious
signal at small angular separations.  Thus we exclude objects closer than
$0.5^{\circ}$ to known Milky Way globular clusters listed in the catalog
of \citet{harris96}.

	Figure \ref{fig:density} shows the density of BHB candidates as a
function of Galactic latitude. The density of BHB candidates increases
from the poles, with a sharp increase beginning around
$|b|\sim45^{\circ}$. We fit a two-component model (shown by the dashed
lines in Figure \ref{fig:density}) to the BHB candidate density
distribution. We use this model to generate random catalogs with the
observed large-scale density distribution.

	We generate 1000 random catalogs with the same area and number of
objects as the BHB candidate catalog.  We run the identical pair-count
program on the random catalogs and the BHB candidate catalog, and then
calculate the angular correlation function.  Figure \ref{fig:angcorr}
shows the two-point angular correlation for a sample of $N=2311$ BHB
candidates with $12.5 < J < 15.5$.  The BHB candidates are located at
($90^{\circ} < l < 270^{\circ}$, $|b|>50^{\circ}$); this region covers
4826 deg$^2$.  Angular bins are 0.2$^{\circ}$ in size, and error bars show
the 1 $\sigma$ scatter about the mean. We choose our final selection
region to avoid the Galactic bulge, the thin and thick disk, the LMC, and
the SMC.  {\it We find no statistically significant structure at any scale
less than $10^{\circ}$ in the high Galactic latitude BHB candidates.} We
also examine narrower ranges in apparent magnitude:  there are no
significant correlations.

	When we include Galactic latitudes below $|b|=50^{\circ}$,
however, we see a systematic rise in the angular correla-

 \includegraphics[width=3.0in]{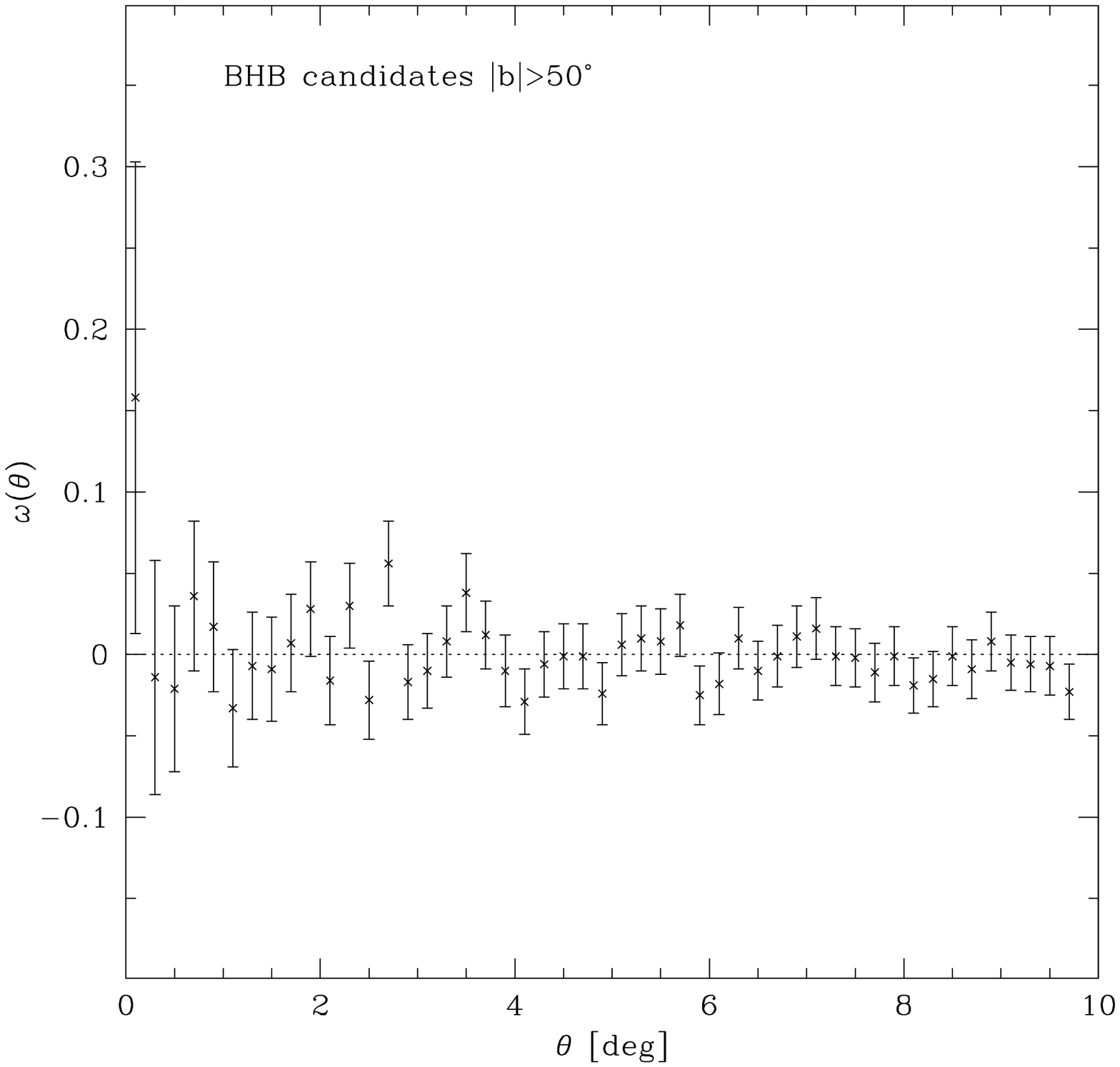}
 \figcaption{ \label{fig:angcorr}
	Two-point angular correlation functions for $N=2311$ BHB
candidates with $12.5 < J_0 < 15.5$ and located ($90^{\circ} < l <
270^{\circ}$, $|b|>50^{\circ}$).  The angular bins are 0.2$^{\circ}$.
Error bars show 1$\sigma$ scatter about the mean.  There is no significant
structure at high Galactic latitudes.}

 \includegraphics[width=3.5in]{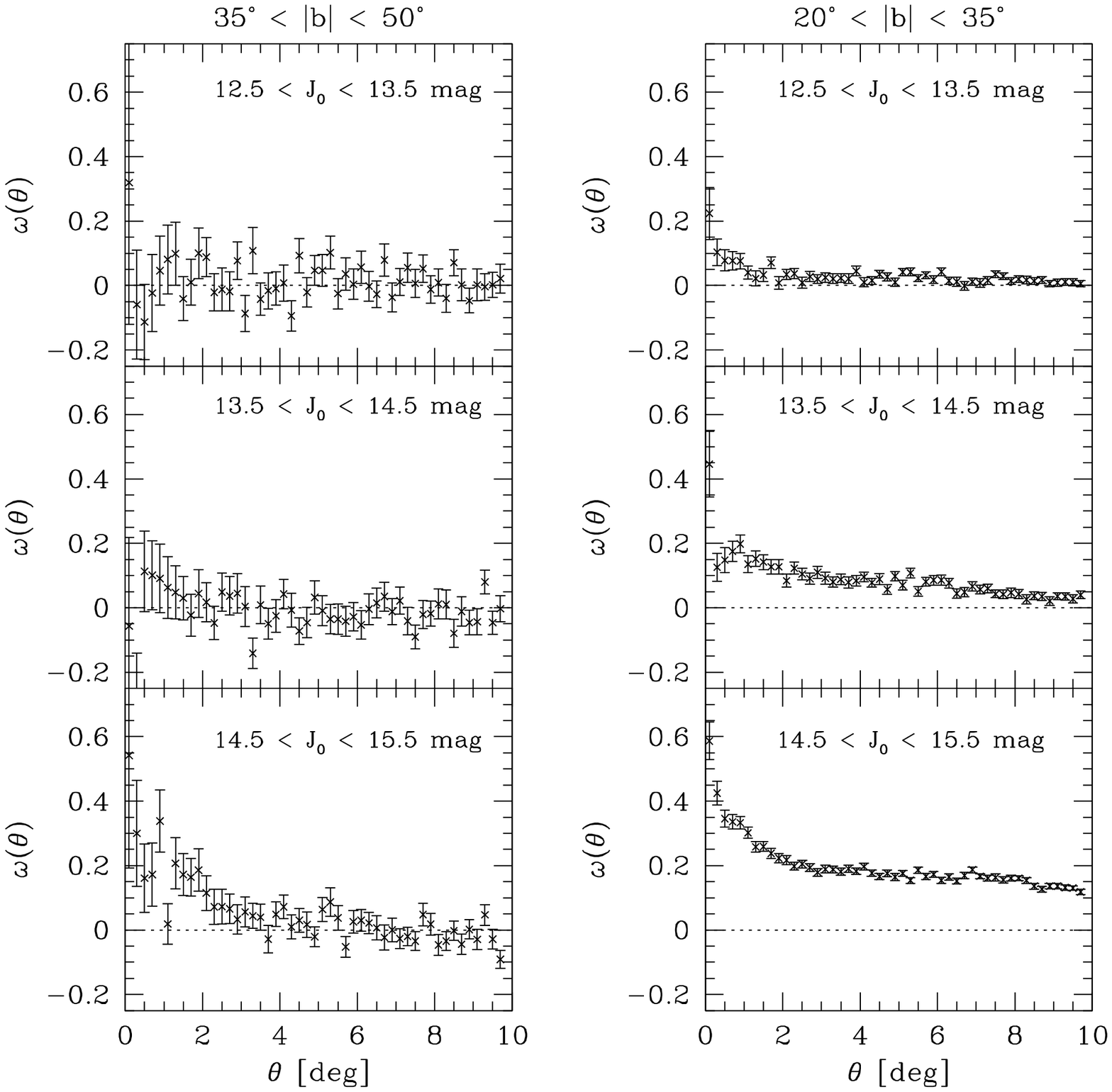}
 \figcaption{ \label{fig:angcorr2}
	Two-point angular correlation functions, calculated in 1
magnitude bins.  The left-hand column is for BHB candidates located
($90^{\circ} < l < 270^{\circ}$, $35<^{\circ}|b|<50^{\circ}$), and the
right-hand column is for BHB candidates located ($90^{\circ} < l <
270^{\circ}$, $20<^{\circ}|b|<35^{\circ}$).  The rising correlation
amplitude at small angular scales appears to be associated with structure
in the thick disk.}

	~

\noindent tion function at $\theta \lesssim 2^{\circ}$ scales. For
example, Figure \ref{fig:angcorr2} shows the two-point angular correlation
calculated for BHB candidates located ($90^{\circ} < l < 270^{\circ}$,
$35^{\circ}<|b|<50^{\circ}$ and $20^{\circ}<|b|<35^{\circ}$). We construct
more pure samples of BHB candidates by decreasing the $(J-H)_0$ color
limit to 0.05, and we still see the same systematic rise in the
small-scale amplitude of the angular correlation at low Galactic
latitudes.

	Table \ref{tab:prop} summarizes the properties of the BHB
candidates samples in Figure \ref{fig:angcorr2}. We re-iterate that the
range of distances $d_{\sun}$ in Table \ref{tab:prop} are
approximate distances that assume $M_J=+0.8$. The range of distances above
the Galactic plane $z$ are calculated using the median Galactic latitude
$b$ in a given sample. When calculating the angular correlation functions,
we fit the density distribution of BHB candidates for the appropriate
magnitude range. Note that the magnitude ranges in Figure
\ref{fig:angcorr2} are identical to the magnitude ranges we use in the
all-sky maps (Figure \ref{fig:skymaps}).

	From inspection of the all-sky maps (Figure \ref{fig:skymaps}), it
is clear that we should expect to find significant correlation at the
lowest Galactic latitudes.  The clump of objects located at $(l,b) \simeq
(170^{\circ}, 35^{\circ})$ in Figure \ref{fig:skymaps}, bottom panel,
causes the strongly increasing angular correlation for $\theta<2^{\circ}$
scales in the lower right panel of Figure \ref{fig:angcorr2}.  Large
angular scale structure near the Galactic plane (structure with
$\theta>10^{\circ}$) likely causes the systematic offset of the
correlation function above zero in the right-hand column of Figure
\ref{fig:angcorr2}. The sharply increasing density gradient near the
Galactic plane may also cause a systematic offset in the correlation
function, though when we adjust our density profile fit near the Galactic
plane, the offset in correlation changes only a small amount.  The angular
correlation at intermediate latitudes $35^{\circ}<|b|<50^{\circ}$, where
the density profile appears quite smooth, remains a surprise.

	One possible explanation for the rising small-scale amplitude in
our correlation analysis at intermediate latitudes is that we are
detecting structure in the \citet{schlegel98} reddening map. The reddening
map is constructed from {\it COBE}/DIRBE data with $0.7^{\circ}$ FWHM
resolution and from {\it IRAS} data with $0.1^{\circ}$ FWHM resolution.
\citet{schlegel98} observe filamentary structure at the smallest scales
resolved by the map. To test whether reddening causes the rise in angular
correlation at intermediate Galactic latitudes, we calculate the angular
correlation for three different magnitude ranges (Figure
\ref{fig:angcorr2}). Because reddening is a foreground effect, it is
intrinsic to all of the BHB candidates and should have the {\it same}
angular scale in different magnitude bins. However, it is clear from
Figure \ref{fig:angcorr2} that the scale of the correlation changes with
apparent magnitude. Thus we rule out residual foreground reddening as the
cause of the rising small-scale amplitude in correlation at low latitudes.

	We associate the rising angular correlation of the BHB candidates
at intermediate Galactic latitudes with some kind of structure in the
thick disk. We note that the angular correlation becomes significant at
the same Galactic latitudes where the stellar density rapidly increases in
Figure \ref{fig:density}. Table \ref{tab:prop} shows that the distance of
these objects above the Galactic plane ranges $1 \lesssim z \lesssim 3$
kpc. The scale height of the thick disk is $\sim$1 kpc \citep[][see their
Table 1]{siegel02}, and the local volume density of the thick disk is
$\sim$50 times that of the halo. Thus stars in the range $1 \lesssim z
\lesssim 3$ kpc are most likely associated with the thick disk and not the
halo.

	Given the estimated range of distances, 0.1$^{\circ}$ and
1$^{\circ}$ angular scales corresponds to {\it physical} scales of
$\sim$10 and $\sim$100 pc. This thick-disk structure could be in the
distribution of stars, for example, moving groups of young A stars from
the thin disk or open clusters that have been deposited by previous dwarf
interactions.  Alternatively, this thick-disk structure could be dark
clouds creating the appearance of a patchy stellar distribution.  We
cannot discriminate among the sources of structure without radial
velocities, but we note that thick-disk structure may be consistent with
expectations from cosmological simulations. \citet{abadi03b} study a
single disk galaxy assembly in a $\Lambda$CDM simulation and find that
60\% of the ``thick disk'' consists of tidal debris from multiple
satellites.  We show below that a simulated star stream can contribute
structure to the correlation function.

	We now compare our results with those of \citet{doinidis89}. The
lack of correlation exhibited by our high Galactic latitude
$|b|>50^{\circ}$ BHB candidates is in clear disagreement with
\citet{doinidis89}. However, the HK Survey fields cover Galactic latitudes
as low as $|b|=35^{\circ}$, with a small number of fields extending to
$|b|=15^{\circ}$. Thus we consider the $35^{\circ}<|b|<50^{\circ}$ BHB
candidate sample a more accurate comparison with \citet{doinidis89}.
Indeed, the bottom left-hand side of Figure \ref{fig:angcorr2} shows the
same features reported by \citet{doinidis89}: a rising correlation towards
smaller angular scales and a significant correlation in the smallest
angular bin. The major difference is that the amplitude of our angular
correlations are approximately half that of \citet{doinidis89}. The
\citet{doinidis89} sample is comprised of roughly 85\% BHB stars. It is
possible that non BHB-stars in our sample dilute our angular correlations,
but we have no expectation that BHB stars cluster more or less strongly
than non-BHB stars. Instead, we believe that the HK Survey fields located
$15^{\circ}<|b|<35^{\circ}$ contribute the excess signal. We note that
these low Galactic latitude fields in the HK Survey are concentrated
around $l=0^{\circ}$ and $l=180^{\circ}$, locations where significant
structure is apparent in Figure \ref{fig:skymaps}. The correlation
function of the $20^{\circ}<|b|<35^{\circ}$ sample of BHB candidates
(Figure \ref{fig:angcorr2} right-hand side) shows correlation at even
higher significance levels than found by \citet{doinidis89}. Thus, it is
likely that structure in the thick disk, present in the HK Survey fields
with $|b|<50^{\circ}$, produces the excess BHB pairs that
\citet{doinidis89} find.

	Finally, we investigate a star stream as a possible source of the
non-zero correlation function in Figure \ref{fig:angcorr2}, and test at
what level we can rule out the presence of a star stream amongst the
high-latitude BHB candidates. The Sagittarius stream in the
\citet{majewski03} plot of 2MASS-selected M giants appears to have a
$\sim5^{\circ}$ FWHM across the southern sky. Using this structure as a
rough guide, we simulate a $5^{\circ}$ wide star stream and insert it into
the high Galactic latitude $|b|>50^{\circ}$ BHB candidate catalog. Figures
\ref{fig:stream}a and \ref{fig:stream}b show the effect of the simulated
stream on the correlation function for surface densities in the stream of
0.5 objects deg$^{-2}$ and 1.0 objects deg$^{-2}$. The underlying BHB
candidate catalog has an average surface density of 0.5 objects
deg$^{-2}$. Figure \ref{fig:stream}a shows that a star stream with the
same density as the BHB candidate catalog results in a detectable increase
in angular correlation at all scales. While a similar systematic offset is
seen in the $20^{\circ}<|b|<35^{\circ}$ BHB candidate sample (Figure
\ref{fig:angcorr2}, right-hand column), there are no visible stream-like
structures in those maps with the appearance of the simulated streams.  
Figure \ref{fig:stream}b more clearly shows the increase in correlation at
small angular scales, and the inflection point at the correct $5^{\circ}$
scale. We use $1^{\circ}$ bins to increase the S/N of the bins,

 \includegraphics[width=3.5in]{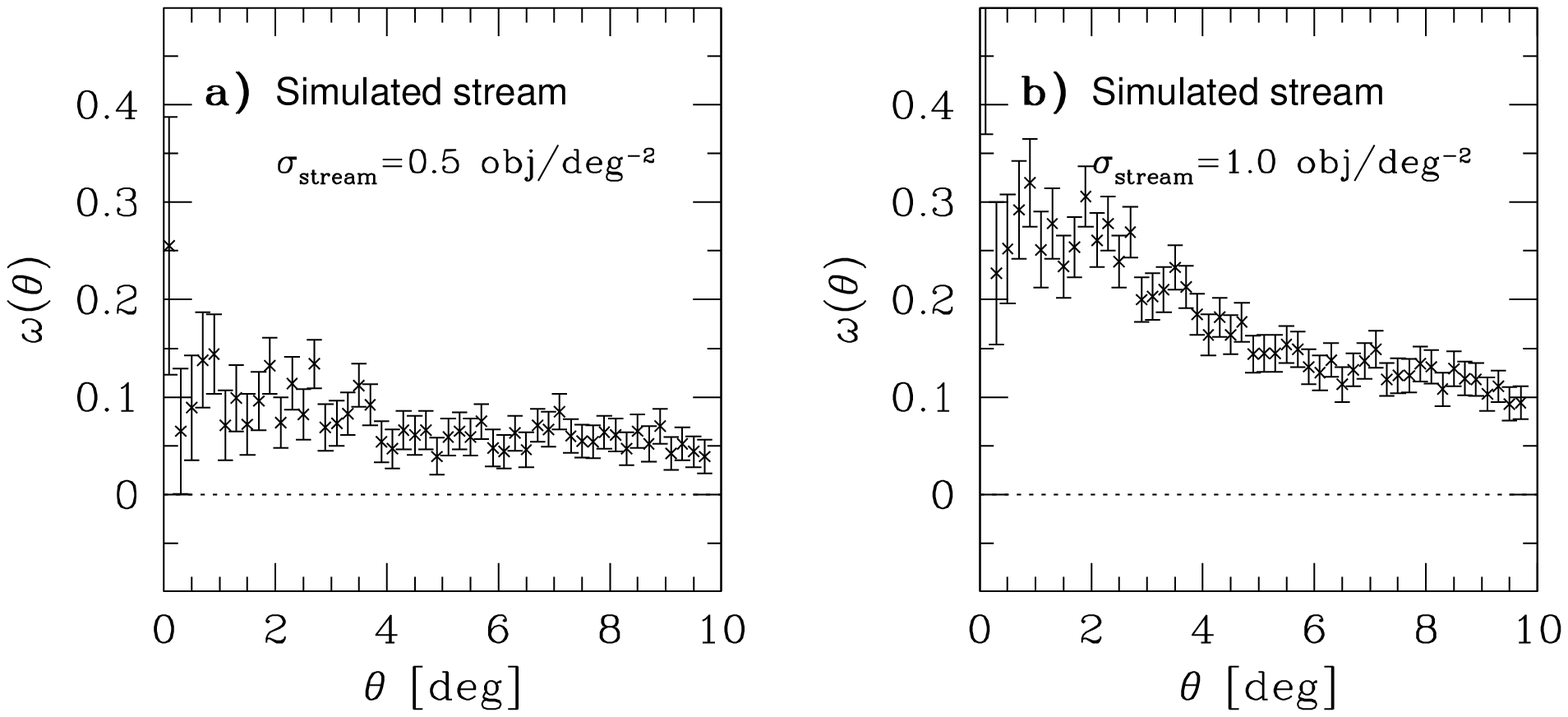}
 \figcaption{ \label{fig:stream}
	Two-point angular correlation functions.  Panels a) and b) show
the effects of a $5^{\circ}$-wide simulated star stream with surface
densities of 0.5 and 1.0 objects deg$^{-2}$, respectively, inserted into
the $|b|>50^{\circ}$ BHB candidates from Figure \ref{fig:angcorr}.}

	~

\noindent and conclude that the high-latitude BHB candidate catalog is
consistent with having no $\sim5^{\circ}$ wide star stream with density
greater than 0.33 objects deg$^{-2}$ at the 95\% confidence level.  We can
place no limit on extremely wide star streams that might cover most of the
sky at the depth of this survey.

\section{CONCLUSIONS AND FUTURE PROSPECTS}

	We use 2MASS near-IR photometry to select BHB candidates for an
all-sky survey. A $12.5<J_0<15.5$ sample of BHB stars ranges
$2<d_{\sun}<9$ kpc, and thus traces the thick disk and inner halo of the
Milky Way. We base the sample selection on the Century Survey Galactic
Halo Project, a survey that provides a complete,
spectroscopically-identified sample of blue stars to the depth of the
2MASS photometry. We investigate the efficacy of 2MASS photometry, and
find that a $-0.20<(J-H)_0<0.10$, $-0.10 < (H-K)_0 < 0.10$ color-selected
sample of stars is 65\% complete for BHB stars, and is composed of 47\%
BHB stars. Increasing the $(J-H)_0$ color limit to $(J-H)_0<0.15$
increases the completeness for BHB stars to 96\%, but reduces the
percentage of bona-fide BHB stars in the sample to 32\%.

	We apply the two-color $-0.20<(J-H)_0<0.10$, $-0.10 < (H-K)_0 <
0.10$ photometric selection to the full 2MASS catalog, and plot the
distribution of the BHB-candidate objects. The 2MASS BHB sample should
include an equivalent number of BHB stars compared to M giants in the
\citet{majewski03} maps.  However, we see no obvious overdensities in the
number counts of the BHB candidates with helio-centric distances
$2<d_{\sun}<9$ kpc. A two-point angular correlation analysis of the BHB
candidates reveals no significant structure at high Galactic latitudes
$|b|>50^{\circ}$. However, we find increasing angular correlation at
$\theta\lesssim1^{\circ}$ for lower Galactic latitudes. This structure may
explain the \citet{doinidis89} result, and suggests that BHB stars in the
thick disk are correlated at scales of 10-100 pc. We propose that clean
samples of the inner halo must be carefully constructed with
$|b|>50^{\circ}$.

	We insert simulated star streams into the data and conclude that
the high Galactic latitude BHB candidates are consistent with having no
$\sim5^{\circ}$ wide star stream with density greater than 0.33 objects
deg$^{-2}$ at the 95\% confidence level. Because stars from disrupted
satellites are spatially well-mixed after a few orbits, the lack of
observed overdensities at high Galactic latitudes suggests there have been
no major accretion events in the inner halo in the last few Gyrs.
\citet{helmi03} show that strong correlations in phase space remain from
past satellite mergers, emphasizing the need for full kinematic
information provided by radial velocities and proper motions.

	In the future we will obtain spectroscopic identifications and
radial velocities for 2MASS-selected BHB candidates as part of the Century
Survey Galactic Halo Project. As our statistics improve, we will be able
to measure an accurate local density of BHB versus other stars. The local
density of BHB stars is a poorly constrained quantity, yet quite important
for the normalization of halo model density profiles. Large numbers of
stars with radial velocities and proper motions (provided by UCAC2 and
SPM) will allow us to carry out a new determination of the Milky Way mass
estimate similar to \citet{sakamoto03}. When the Sloan Digital Sky Survey
is complete, color-selected BHB candidates at helio-centric distances up
to 75-100 kpc will be available to complement the 2MASS-selected inner
halo sample. Similarly deep color-selected samples of BHB candidates are
expected to be available shortly from the GALEX satellite mission (Rhee,
private communication). The combination of the inner-halo sample with mid-
and distant-halo samples will provide a definitive map of the distribution
of the thick-disk/halo BHB populations of the Milky Way.

\acknowledgements

	We thank the anonymous referee for a prompt, insightful, and
constructive report.  This project makes use of data products from the Two
Micron All Sky Survey, which is a joint project of the University of
Massachusetts and the Infrared Processing and Analysis Center/California
Institute of Technology, funded by NASA and the NSF.  This research also
makes use of the SIMBAD database, operated at CDS, Strasbourg, France. TCB
acknowledges partial support for this work from NSF grants AST 00-98508
and AST 00-98549 awarded to Michigan State University.

\appendix
\section{CONSTRUCTION OF THE ALL-SKY MAPS}

It is a non-trivial task to generate the Hammer-Aitoff projection
in Figure \ref{fig:skymaps}, hence the interested reader may want to know
what software packages we used.  The Starbase Data Table software package
\citep{roll96}, available at
\url{http://cfa-www.harvard.edu/\~{}john/starbase/starbase.html}, uses an
ASCII table format and a set of filter programs to work with astronomical
data.  We used Starbase to manipulate the 2MASS catalog, and to perform
our final object selection.  The Generic Mapping Tools \citep{wessel91},
available at \url{http://gmt.soest.hawaii.edu/}, are designed to perform
map projections for the geophysics community.  We use Generic Mapping
Tools to create the equal-area Hammer-Aitoff projections. The Fits Users
Need Tools package \citep{mandel01}, available at
\url{http://hea-www.harvard.edu/RD/funtools/}, provides simplified access
to fits images and binary tables for astronomical data.  We use the Fits
Users Need Tools to convert the Hammer-Aitoff projections to fits images.

Our two-color-selected catalog of candidate BHB stars in the
2MASS point source catalog is available to the community at
\url{http://tdc-www.harvard.edu/chss/}.



\begin{deluxetable}{lcccc}
\tablewidth{0pt}
\tablecaption{Properties of the Correlation Function Samples\label{tab:prop}}
\tablecolumns{5}
\tablehead{
	\colhead{$b$ range,~~~~~~~~~} &
	\colhead{$N$} &
	\colhead{Area} &
	\colhead{$d_{\sun}$\tablenotemark{a}} &
	\colhead{$z$\tablenotemark{a}} \\
	\colhead{mag range} & \colhead{} &
	\colhead{deg$^2$} & \colhead{kpc} & \colhead{kpc}
}
	\startdata
$|b|>50^{\circ}$	& & & & \\
$~~12.5<J_0<15.5$	& 2311 & 4826 & 2.2 - 8.7 & 1.9 - 7.6 \\

$35<|b|<50^{\circ}$	& & & & \\
$~~12.5<J_0<13.5$	&  860 & 3970 & 2.2 - 3.5 & 1.4 - 2.3 \\
$~~13.5<J_0<14.5$	&  914 & 3970 & 3.5 - 5.5 & 2.3 - 3.6 \\
$~~14.5<J_0<15.5$	& 1180 & 3970 & 5.5 - 8.7 & 3.5 - 5.7 \\

$20<|b|<35^{\circ}$	& & & & \\
$~~12.5<J_0<13.5$	& 3814 & 4776 & 2.2 - 3.5 & 0.9 - 1.4 \\
$~~13.5<J_0<14.5$	& 3418 & 4776 & 3.5 - 5.5 & 1.4 - 2.3 \\
$~~14.5<J_0<15.5$	& 5693 & 4776 & 5.5 - 8.7 & 2.3 - 3.6 \\
	\enddata
 \tablenotetext{a}{Distance estimates assume $M_J=+0.8$ as explained in
\S \ref{sec:maps}.}
 \end{deluxetable}

\end{document}